\documentclass[letter,twocolumn]{jpsj3}

\newcommand{\lsim} 
 {\ \raise.35ex\hbox{$<$}\kern-0.75em\lower.5ex\hbox{$\sim$}\ }
\newcommand{\gsim}
 {\ \raise.35ex\hbox{$>$}\kern-0.75em\lower.5ex\hbox{$\sim$}\ }

\usepackage{graphicx}
\usepackage{dcolumn}
\usepackage{bm}
\usepackage{amsmath}
\usepackage{times}
\usepackage[usenames]{color}

\title{Collective Charge Excitation in a Dimer Mott Insulating System}  
\author{Makoto~Naka$^1$ and Sumio Ishihara$^{1,2}$}
\inst{
$^1$Department of Physics, Tohoku University, Sendai 980-8578, Japan \\
$^2$Core Research for Evolutional Science and Technology (CREST), Sendai 980-8578, Japan}
\abst{  
Charge dynamics in a dimer Mott insulating system, where a non-polar dimer-Mott (DM) phase and a  polar charge-ordered (CO) phase compete with each other, are studied. 
In particular, collective charge excitations are analyzed in the three different models where the internal-degree of freedom in a dimer is taken into account.  
Collective charge excitation exists both in the non-polar DM phase and the polar CO phase, and softens in the phase boundary. 
This mode is observable by the optical conductivity spectra where the light polarization is parallel to the electric polarization in the polar CO phase. 
Connections between the present theory and the recent experimental results in $\kappa$-(BEDT-TTF)$_2$Cu$_2$(CN)$_3$ are discussed. 
}

\kword{dimer-Mott insulator, ferroelectricity, charge order, optical conductivity, collective mode}

\begin{document}
\maketitle

Exotic ferroelectricity beyond the conventional displacive-type and order-disorder-type classifications has been recognized as one of the central issues in recent condensed matter physics. Multiferroics, i.e. coexistence of ferroelectricity, ferroelasticity and ferromagnetism, are one of the examples. Non-collinear spin orders in Mott insulators break the space-inversion symmetry through the symmetric- and antisymmetric-exchange interactions. In this sense, the series of materials is classified as a spin-driven ferroelectricity. Another example is seen in the so-called quarter-filled system. 
At low temperature, an electronic charge order occurs so as to break the space-inversion symmetry. This is termed charge-driven ferroelectricity or electronic ferroelectricity~\cite{khomskii, ishihara}. 
Several transition-metal oxides and organic compounds have been studied intensively and extensively as candidates of the electronic ferroelectricity. 
Characteristics in electronic ferroelectricity are dielectric fluctuation. 
Since the electron mass is much smaller than the ion mass and the proton mass, large dielectric fluctuation due to electronic charge dynamics is expected to play a crucial role on the ferroelectricity. 

A layered organic compound $\kappa$-(BEDT-TTF)$_2$Cu$_2$(CN)$_3$ and related materials are recognized as candidates of electronic ferroelectric materials. 
A crystal lattice consists of cation and anion layers stacked along the $z$ axis. 
In a cation layer, two BEDT-TTF (ET) molecules form a dimer where bonding and antibonding orbitals are constructed from the ET molecular orbitals (MO). 
Since one hole exists per dimer in the chemical formula, the antibonding MO is occupied by one hole. 
Thus, this system is identified as a Mott insulator, termed a dimer-Mott (DM) insulator, in the strong dimerizaion limit~\cite{kino,seo}.  
No long-range spin order down to few mK suggested experimentally has attracted a number of researchs in the viewpoint of realization of spin-liquid state~\cite{shimizu,syamashita,myamashita}. 
Recently, an anomalous enhancement in the dielectric constant is discovered around 30K~\cite{jawad}. This observation triggers reconsideration of the DM insulator picture and suggests a possibility of the polar charge distribution inside dimers and electronic ferroelectricity~\cite{naka_kappa,hotta,gomi,dayal}. 

In this Letter, charge dynamics in a DM insulating system are examined. 
In particular, the collective excitations in the non-polar DM phase and the polar charge ordered (CO) phase are focused on. 
Collective excitation is the central concept not only in ferroelectricity but also in other symmetry broken state~\cite{blinc,degennes,tokunaga}. The soft-phonon mode in the displacive-type ferroelectricity, the soft-proton mode in the hydrogen-bond type ferroelecticity, and the magnon induced by electric field in multiferroics are typical examples. 
We propose that the collective charge excitation also exists in the DM insulating system, where the non-polar DM phase and the polar CO phase compete with each other. 
This mode represents the polarization flip in the polar CO phase and the intra-dimer bonding-antibonding excitation in the DM phase, and softens in the phase boundary. 
This soft collective mode is observable by the optical conductivity spectra where the light polarization is parallel to the electric polarization in the polar CO phase. 
Connection between the present theory and the recent experimental results in $\kappa$-(ET)$_2$Cu$_2$(CN)$_3$~\cite{itoh, nakaya} are discussed. 

To examine the charge dynamics without ambiguity, we analyze the three different models, the extended-Hubbard model (EHM), the spin-less $Vt$ model (VtM) and the pseudo-spin model (PSM), where two molecules denoted by $a$ and $b$ are introduced in each dimer located in a crystal lattice.  
The EHM is defined as 
\begin{align}
{\cal H}_{\rm EH}
&=U \sum_{i \mu s} n_{i \mu \uparrow} n_{i \mu \downarrow} 
\nonumber \\
&+ t_A \sum_{i s} \left ( c_{i a s}^\dagger c_{i b s}^{}+ H.c. \right ) + V_A \sum_{i} n_{i a} n_{i b}, \nonumber \\
&+ \sum_{\langle ij \rangle \mu \mu' s}
t_{ij}^{\mu \mu'} 
\left ( c_{i \mu s}^\dagger c_{j \mu' s}+H.c. \right )  + \sum_{\langle ij \rangle \mu \mu'} V_{i j}^{\mu \mu'} n_{i \mu } n_{j \mu'}, 
\label{eq:ehm} 
\end{align}
where $c_{i \mu s}$ is an annihilation operator for a hole at the $i$-th dimer with spin $s(=\uparrow, \downarrow)$ and molecule $\mu(=a, b)$, and $n_{i \mu}(\equiv \sum_{s} n_{i \mu s}= \sum_s c_{i \mu s}^\dagger c_{i \mu s})$ is a number operator. 
We introduce the intra-molecule Coulomb interaction $(U)$, 
the intra-dimer transfer $(t_A)$, the intra-dimer Coulomb interaction $(V_A)$, 
the inter-dimer transfer $(t_{ij}^{\mu \mu'})$, and the inter-dimer Coulomb interactions $(V_{i j}^{\mu \mu'})$.
When we focus on the charge dynamics where 
the excitation energy is less than the effective intra-dimer Coulomb interaction $U_{\rm eff} = [ U+V_A+4t_A - \sqrt{(U-V_A)^2+16t_A^2} ]/2$, 
the spin-less VtM is valuable as an effective model. This is defined by 
\begin{align}
{\cal H}_{Vt}
&= t_A \sum_{i} \left ( f_{i a}^\dagger f_{i b}^{}+ H.c. \right ) 
+V_A \sum_{i} n'_{i a} n'_{i b} \nonumber \nonumber \\
&+\sum_{\langle ij \rangle \mu \mu'} t_{ij}^{\mu \mu'} \left ( f_{i \mu}^\dagger f_{j \mu'}+H.c. \right ) 
+ \sum_{\langle ij \rangle \mu \mu'} V_{i j}^{\mu \mu'} n'_{i \mu} n'_{j \mu'} , 
\label{eq:hvt}
\end{align}
where $f_{i \mu}$ is an annihilation operator for a spin-less fermion at the $i$-th dimer with molecule $\mu$, and $n'_{i \mu}(\equiv f_{i \mu }^\dagger f_{i \mu })$ is a number operator.
The transfers and the Coulomb interactions for fermion in the VtM are defined in the same way with those in the EHM. 

When the fermion number in each dimer is fixed to be one, the charge state in a dimer is described by the pseudo-spin (PS) operator with an amplitude of $1/2$. 
This is defined as 
${\bm Q}_{i}=(1/2) \sum_{\mu \mu'} {\hat f}_{i \mu}^\dagger {\bm \sigma}_{\mu \mu'} {\hat f}_{i \mu'}$ 
with the Pauli matrices $\bm \sigma$ and $\hat f_{i \mu}=\sum_{\nu} U_{\mu \nu} f_{i \nu}$ where $U \equiv (1/\sqrt{2}) ( \sigma^z + \sigma^x )$. 
The eigen states for $Q^x_i$ with the eigen values of $1/2$ and $-1/2$ are the charge polarized states $|a \rangle \equiv f_{ia}^\dagger|0 \rangle$ and $|b \rangle \equiv f^\dagger_{ib}|0 \rangle$, respectively, 
and those for $Q^z_i$ with $1/2$ and $-1/2$ are the bonding state  
$|\beta \rangle=(|a \rangle + |b \rangle)/\sqrt{2}$ 
and the antibonding state $|\alpha \rangle=(|a \rangle - |b \rangle)/\sqrt{2}$, respectively. 
When we assume in the VtM that the inter-dimer transfers and Coulomb interactions are smaller than 
the intra-dimer Coulomb interactions, we obtain the low-energy effective Hamiltonian up to the orders of $O({\cal H}_t^2)$ and $O({\cal H}_V^1)$ as  
\begin{align}
{\cal H}_{\rm PS}=
2t_A \sum_{i} Q^z_{i} + \sum_{\langle ij \rangle} W_{ij} Q_i^x Q_j^x 
+{\cal H}_J , 
\label{eq:tising}
\end{align}
where the first and second terms originate from the first and last terms in Eq.~(\ref{eq:hvt}), respectively, and $W_{ij}(\equiv V_{ij}^{aa}+V_{ij}^{bb}-V_{ij}^{ab}-V_{ij}^{ba})$ is the effective inter-dimer Coulomb interaction. 
The third term represents the exchange interactions, 
which are given by 
\begin{align}
{\cal H}_J &= \sum_{\langle ij \rangle  (\gamma, \gamma')=(\pm)} \left ( 
J_{ij}^{\gamma \gamma'} m_{i}^{ \gamma} m_{j}^{ \gamma'} 
+  K_{ij}^{\gamma \gamma'} Q_{i}^{ \gamma} Q_{i}^{ \gamma'} 
\right )
\nonumber \\
&+ \sum_{\langle ij \rangle \gamma=(\pm)} 
\left( I_{ij}^{x \gamma} Q_{i}^{x} m_{j}^{\gamma} + I_{ij}^{\gamma x} m_{i}^{ \gamma} Q_{j}^{x} \right), 
\label{eq:hj}
\end{align}
where we introduce $Q_{i}^{\pm} = Q_{i}^{x} \pm i Q_{i}^{y}$ and $m_{i}^{\pm} = \frac{1}{2} \pm Q_{i}^{z}$. 
The exchange constants are given in Ref.~\cite{exchange}. 
We note that a related model in a one-dimensional system is analyzed in Ref.~\cite{tsuchiizu}. 

\begin{figure}[t]
\begin{center}
\includegraphics[width=\columnwidth,clip]{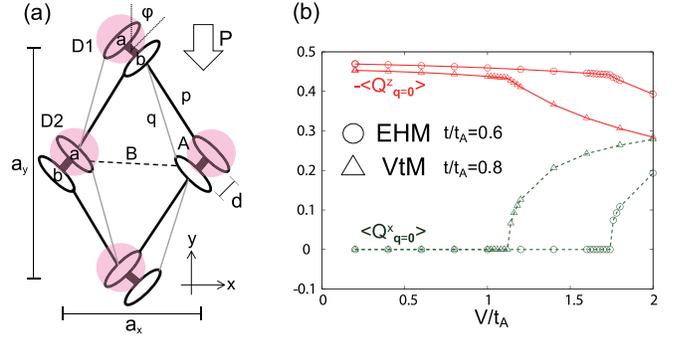}
\end{center}
\vspace{-0.2cm}
\caption{ 
(a) A schematic lattice structure for $\kappa$-(ET)$_2$Cu$_2$(CN)$_3$. 
Ellipses represent the ET molecules, and two molecules in a dimer are indicated by $a$ and $b$. 
Black and gray lines represent the transfers and the Coulomb interactions introduced in the EHM and the VtM, and symbols $A$, $B$, $p$ and $q$ correspond to the subscripts in $t$ and $V$ in the Hamiltonians (see text). 
Two inequivalent dimers in a unit-cell are denoted by $D1$ and $D2$. 
We take lattice constants $a_{x}=1$ and $a_{y}=\sqrt{3}$, a distance between the two ET molecules in a dimer $d=0.5$, and the angle of the molecular plane $\phi = \pi/4$. 
Shaded circles and a thick arrow represent schematic hole distributions and a direction of the electric polarization in the polar CO phase, respectively.
(b) The DM order parameter $-\langle Q^z_{{\bm q}=0}\rangle$ (soild lines) 
and the CO parameter $\langle Q^x_{{\bm q}=0}\rangle$ (dash lines) for the 8-dimer cluster of the EHM, 
and those for the 12-dimer cluster in the VtM. 
Parameter values are chosen to be $U = 8$, $V_{A} = 2$ for the EHM and 
$V_{A} = 2$ for the VtM.  
}
\label{fig:lattice}
\vspace{-0.4cm}
\end{figure}

The EHM and the VtM are analyzed by the combined method of 
the exact-diagonalization based on the Lanczos algorithm, 
and the mean-field approximation. 
We introduce the mean-fields, which act on the edge-sites of the cluster,  
by decoupling the inter-dimer Coulomb interactions, such as  
$n_{i \mu} n_{j \mu'} \rightarrow \langle n_{i \mu} \rangle n_{j \mu'} + n_{i \mu} \langle n_{j \mu'} \rangle - \langle n_{i \mu} \rangle \langle n_{j \mu'} \rangle $. 
The mean fields are determined to be consistent with the electronic state inside the cluster. 
The charge dynamics are examined by calculating the optical absorption spectra defined by 
${\alpha_{\xi} ( \omega ) } = -(e^{2}/N) 
{\rm Im}  \langle 0 | j_{\xi} ( \omega-{\cal H} + E_{0}+ i \eta )^{-1} j_{\xi} | 0 \rangle$, 
and the PS dynamical correlation functions defined by 
${N_{\xi}^{(\pm)} ( {\bm q}, \omega ) } = 
-{\rm Im}  \langle 0 | L^{\xi (\pm)}_ {\bm q} ( \omega-{\cal H} + E_{0} + i \eta )^{-1} L^{\xi (\pm)}_{- {\bm q}} | 0 \rangle$, 
where $\xi$ takes a two-dimensional Cartesian coordinate, $|0\rangle$ and $E_{0}$ are the ground-state wave-function and energy, respectively, $\eta$ is an infinite decimal constant, and $N$ is the number of dimers in a cluster. 
We define the current operators 
${\bm j} = i \sum_{\langle ij \rangle \mu \mu' s} t_{ij}^{\mu \mu'} 
({\bm R}_{i \mu} - {\bm R}_{j \mu'})
( c_{i \mu s}^{\dagger} c_{j \mu' s} - c_{j \mu' s}^{\dagger} c_{i \mu s}  )$ 
for the EHM and 
${\bm j} = i \sum_{\langle ij \rangle \mu \mu'} t_{ij}^{\mu \mu'}
({\bm R}_{i \mu} - {\bm R}_{j \mu'})
( f_{i \mu}^{\dagger} f_{j \mu'} - f_{j \mu'}^{\dagger} f_{i \mu} )$ 
for the VtM,  
and 
$L^{\xi (\pm)}_{\bm q} = (1/N) 
\left ( \sum_{i \in D1} \pm \sum_{i \in D2} \right )
Q^{\xi}_{i}\exp ( -i {\bm q} \cdot {\bm R}_{i} )$ 
where $\sum_{i \in D1 (D2)}$ represents a summation for the site $i$ which belongs to the sublattice ${D1} (D2)$ 
and ${\bm R}_{i}=({\bm R}_{ia}+{\bm R}_{ib})/2$. 
A schematic arrangement of molecules and interactions are shown in Fig.~\ref{fig:lattice}(a). 
By taking into account the relations of $t_{A} \gg t_{B} \cong t_{p} \gg t_{q}$
in $\kappa$-(ET)$_2$Cu$_2$(CN)$_3$,  we set the intra-dimer transfer $t_{A}$ as a unit of energy, and $t_{q}=0$ and $t_{p} = t_{B} (\equiv t)$, for simplicity. 
We consider the $1/r$-type dependence for the Coulomb interactions, 
and set $V_{q} = 0.7V_p$ and $V_{B} = 0.7V_p$ for the EHM, and $V_{q} = 0.9V_p$ and $V_{B} = 0.8V_p$ for the VtM as a unit of $V_{p} (\equiv V)$.

First we show the ground state properties. 
In Fig.~\ref{fig:lattice}(b), 
we show the DM order parameter $-\langle Q^z_{{\bm q}=0}\rangle$ and 
the polar CO parameter $\langle Q^x_{{\bm q}=0}\rangle$ in the EHM. 
With increasing $V$,  
there is a critical $V(\equiv V_c)$ where 
$-\langle Q^z_{{\bm q}=0}\rangle$ decreases and $\langle Q^x_{{\bm q}=0} \rangle$ becomes finite from zero. 
The phase transition from the DM phase to the polar CO phase occurs at $V=V_c$. 
The charge configuration in the polar CO phase is illustrated in Fig~\ref{fig:lattice} (a) where 
the electric polarization appears along the $y$ direction. 
Numerical results in the VtM [see Fig.~\ref{fig:lattice}(b)] are qualitatively the same with those in the EHM.

\begin{figure}[t]
\begin{center}
\includegraphics[width=\columnwidth,clip]{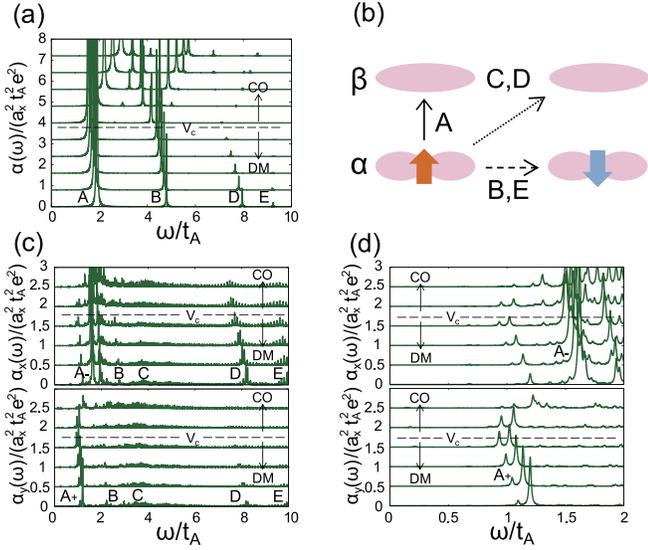}
\end{center}
\vspace{-0.2cm}
\caption{
(a) The optical absorption spectra for the EHM of a 2-dimer cluster. 
Parameter values of $V$ are changed from $0.4$ to $4$ from the bottom to the top, 
and the DM and polar CO phases are realized in $V<V_C=2.39$ and $V>V_C$, respectively.
Other parameters are chosen to be $U = 6$, $V_{A} =4$, $t = 0.5$ and $\eta=0.01$. 
(b) Schematic intra-dimer and inter-dimer charge excitations. 
Symbols $\alpha$ and $\beta$ represent antibonding and bonding orbitals, respectively, and 
A, B, C, D and E correspond to the peaks in (a), (c) and (d) (see text). 
(c) The optical absorption spectra $\alpha_{x} \left( \omega \right)$ (the upper panel) and $\alpha_{y} \left( \omega \right)$ (the lower panel) for the EHM of an 8-dimer cluster.
Parameter values of $V$ are changed from $1$ to $2$ from the bottom to the top, 
and other parameters are chosen to be  $U = 8$, $V_{A}=2$, $t = 0.6$ and $\eta=0.01$. 
(d) The low energy parts of $\alpha_{x,y} \left( \omega \right)$ in (c). 
}
\label{fig:EHM}
\vspace{-0.4cm}
\end{figure}

Now we present the charge dynamics in the EHM. 
We begin with a simple two-dimer cluster. 
The optical absorption spectra for various $V$ are shown in Fig.~\ref{fig:EHM}(a). 
In the DM phase, mainly four peaks, termed A, B, D and E, 
appear for small $V$. 
The lowest-energy peak A,
shows a softening around the phase boundary. 
In the polar CO phase, the peak B 
splits, and intensities of the peak D and E are reduced. 
The peaks B, D, and E 
are identified as the inter-dimer charge excitations, described as 
${\cal D}_i^1{\cal D}_j^1 \rightarrow {\cal D}_i^2{\cal D}_j^0$, 
where ${\cal D}_i^n$ represents a state at the $i$-th dimer with $n$ holes. 
On the other hand, the peak 
A is the intra-dimer charge excitation, 
${\cal D}_i^1{\cal D}_j^1 \rightarrow {\cal D}_i^{1\ast} {\cal D}_j^{1 }$, where ${\cal D}_i^{n \ast}$ represents the excited state of ${\cal D}_i^{n }$. 
We note that the peak A is connected to the collective excitation of the present interest. 

The ${\cal D}^2$ states in a single dimer are classified as the triplet states, 
$|T \rangle 
= \{ |\alpha_\uparrow \beta_\uparrow \rangle, |\alpha_\downarrow \beta_\downarrow \rangle, (|\alpha_\uparrow \beta_\downarrow \rangle+|\alpha_\downarrow \beta_\uparrow \rangle)/\sqrt{2} \} $, 
the singlet state, $|S \rangle =(|\alpha_\uparrow \beta_\downarrow \rangle -|\alpha_\downarrow \beta_\uparrow \rangle) /\sqrt{2} $, 
and the states where two holes occupy the same MO as, 
$|D_\pm \rangle=C_\pm| \alpha_\uparrow \alpha_\downarrow \rangle \pm C_\mp|\beta_\uparrow \beta_\downarrow \rangle$. 
Eigen energies are
$E_T=V_A$, $E_{S}=U$, and $E_{D\pm}=[ U+V_A \pm \sqrt{(U-V_A)^2+16t_A^2} ]/2$, respectively, and coefficients are given by $C_-/C_+=(U-V_A)/[2E_{D+}-4t_A-(U+V_A)]$. 
The peaks 
B, D, and E originate from the inter-dimer excitations 
where the final ${\cal D}^2$ states are $|D_-\rangle$, $| S \rangle$ and $| D_+ \rangle$, respectively. 
Since the two holes occupy the same MO and the different two MO in $|D_\pm \rangle$ and $|S \rangle$, respectively, 
the peaks 
B and E are termed the Hubbard excitations, and the peak 
D is termed the inter-MO excitation [see Fig.~\ref{fig:EHM}(b)]. 
%

The results in a 8-dimer cluster for the EHM are presented in Figs.~\ref{fig:EHM}(c) and (d).
We identify the peaks A-E in the same way with the results in the two-dimer cluster, 
although multiple-peak structures are induced by the lattice effect. 
Two-excitation modes, termed A$_{+}$ and A$_{-}$, for the intra-dimer excitation 
appear. 
The lower- (higher-) energy mode is seen for the $y (x)$ polarization, and 
the lower-energy mode shows softening around the phase boundary.

\begin{figure}[t]
\begin{center}
\includegraphics[width=\columnwidth,clip]{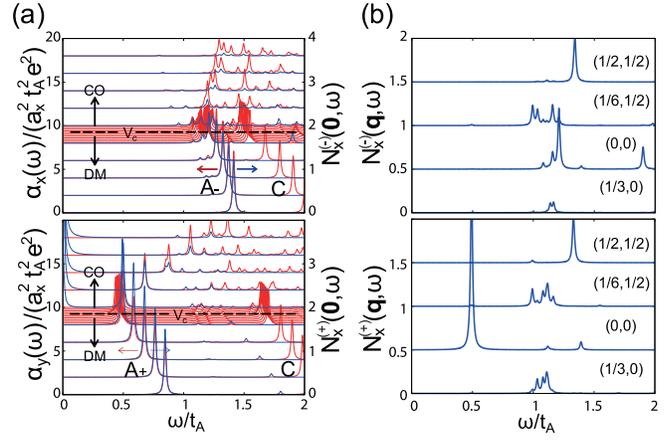}
\end{center}
\vspace{-0.2cm}
\caption{(Color online)
(a) The optical absorption spectra $\alpha_{x} \left( \omega \right)$ (the upper panel) and $\alpha_{y} \left( \omega \right)$ (the lower panel) for the VtM of a 12-dimer cluster (red lines). 
Parameter values of $V$ are changed from $0.2$ to $2$ from the bottom to the top, 
and the DM and CO phases are realized in $V<V_C=1.13$ and $V>V_C$, respectively.
Other parameters are chosen to be $V_{A} = 2$, $t = 0.8$ and $\eta=0.01$. 
The dynamical PS correlation function $N_x^{(+)}({\bm q}={\bm 0}, \omega)$ are also plotted by blue lines for comparison. 
(b) The dynamical PS correlation functions  $N_x^{(\pm)}({\bm q}, \omega)$ 
in the energy-momentum plane near the phase boundary ($V = 1$). 
Other parameter values are the same with those in (a). 
}
\label{fig:vt}
\vspace{-0.4cm}
\end{figure}
%
Detailed low-energy excitations are examined in larger clusters of the VtM.  
The optical absorption spectra in a 12-dimer cluster are shown by red lines in Fig.~\ref{fig:vt}(a). 
Spectra around $1.1 \lesssim \omega \lesssim 1.4$ in $\alpha_{x} (\omega)$ and $0.4 \lesssim \omega \lesssim 1.2$ in $\alpha_{y} ( \omega )$ are attributed to the intra-dimer charge excitations, corresponding to the peak 
A$_{-}$ and A$_{+}$, respectively, in Fig.~\ref{fig:EHM}(c) and (d). 
Remarkable polarization dependence and softening around the phase boundary are observed as seen in the results for the EHM. 
The dynamical PS correlation functions, $N_{x}^{(\pm)}({\bm q}={\bm 0}, \omega)$ are compared with $\alpha_{x, y} (\omega)$ in Fig.~\ref{fig:vt}(a). 
The peak positions in $N_{x}^{(-)} ({\bm 0}, \omega )$ and $N_{x}^{(+)} ({\bm 0}, \omega )$ almost coincide with those in the low-energy modes of $\alpha_{x} (\omega)$ and $\alpha_{y} ( \omega)$, respectively. 
Since the off-diagonal PS operators describe the electronic-state change inside a dimer, 
this coincidence implies that the lower-energy modes in $\alpha_\xi(\omega)$ are attributed to the intra-dimer charge excitations. 
In the CO phase, the intensity of $N_{x}^{(\pm)} ({\bm q}={\bm 0}, \omega )$ decreases with increasing $V$, because the direction of PS is changed from $Q^z$ to $Q^x$.  
The peaks at $\omega = 0$ in $N_{x}^{(+)} ({\bm q}={\bm 0}, \omega )$ are the superlattice Bragg peaks due to the CO. 
The momentum dependence of $N_{x}^{(\pm)} ({\bm q}, \omega )$ 
near the phase boundary is presented in Fig.~\ref{fig:vt}(b). 
A clear dispersion seen in the low-energy mode suggests that this mode is a collective charge excitation. 

\begin{figure}[t]
\begin{center}
\includegraphics[width=1.0\columnwidth,clip]{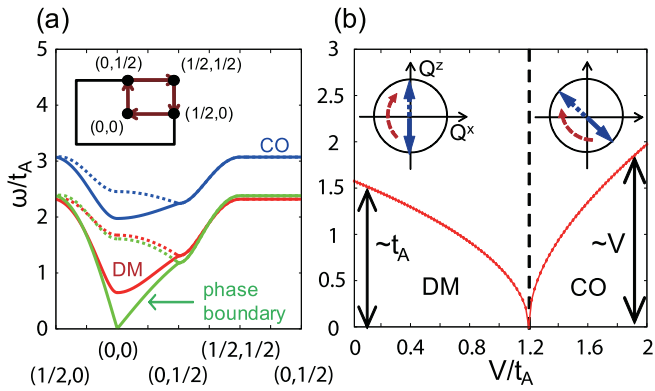}
\end{center}
\vspace{-0.2cm}
\caption{(Color online)
(a) The energy dispersions of the PS waves.  
Bold and broken lines are for $\epsilon_{\bm q}^{\left( + \right)}$ and $\epsilon_{\bm q}^{\left( - \right)}$, respectively. 
Red, green and blue lines show the results in the DM phase ($V=1$), at the phase boundary ($V=1.2 \equiv V_c$), and in the CO phase ($V=2$), respectively. 
The inset shows the first Brillouin zone for the rectangle unit cell shown in Fig.~\ref{fig:lattice}(a), 
and arrows indicate the wave vector trajectories. 
Parameter values are chosen to be $V_{A}=3$ and $t=0.5$. 
(b) The energy of the acoustic PS wave, $\epsilon_{{\bm q}={\bm 0}}^{( +)}$, in the DM and CO phases. 
Schematic PS motions for the collective mode are shown in the inset. 
}
\label{fig:sw}
\vspace{-0.4cm}
\end{figure}

To examine the low-energy excitations without the finite-size effect, 
the PSM is analyzed by the spin-wave approximation. 
The PS order parameters are adopted as 
$\langle Q^z_{\bm q} \rangle = - \frac{1}{2} \delta_{{\bm q}={\bm 0}}$ and $\langle Q^x_{\bm q} \rangle = 0$ for the DM phase, and  
$\langle Q^z_{\bm q} \rangle = - \frac{1}{2} \delta_{{\bm q}={\bm 0}} \cos \theta$ and $\langle Q^x_{\bm q} \rangle = \frac{1}{2} \delta_{{\bm q}={\bm 0}} \sin \theta$ for the polar CO phase, where the angle $\theta$ is determined to minimize the ground state energy. 
%
%
By applying the Holstein-Primakoff transformation and the Bogoliubov transformation to the PS operators, 
two kinds of bosons originating from the two-inequivalent dimers in a unit cell are introduced. 
The dispersion relations for the two modes are obtained as 
$\epsilon_{\bm q}^{\left( \pm \right)} = \sqrt{ ( J_{A {\bm q}} \pm J_{B {\bm q}} )^{2} - \{ J_{C  {\bm q}} \pm J_{D {\bm q} } )^{2}} 
$, 
where $J_{A {\bm q}}$, $J_{B {\bm q}}$, $J_{C {\bm q}}$ and $J_{D {\bm q}}$ are given in Ref.~\cite{sw}. 
In this scheme, the dynamical PS correlation functions are given by 
$N_{x}^{(\pm)} ({\bm q}, \omega ) 
= \frac{1}{4 N} [ \cosh \theta_{\bm q}^{( \pm )} - \sinh \theta_{\bm q}^{ \pm } ] 
\delta  ( \omega - \epsilon_{\bm q}^{ \pm }  ), 
$
where    
$\cosh \theta_{\bm q}^{( \pm )} = ( J_{A {\bm q}} \pm J_{B {\bm q}} ) / \epsilon_{\bm q}^{( \pm )}$ and 
$\sinh \theta_{\bm q}^{( \pm )} = ( J_{C {\bm q}} \pm J_{D {\bm q}} ) / \epsilon_{\bm q}^{( \pm )}$.  
It is shown that $N_{x}^{(\pm)} ({\bm q}, \omega )$ takes peaks at $\varepsilon^{(\pm)}_{\bm q}$. 
The $\epsilon_{\bm q}^{\left( + \right)}$ and $\epsilon_{\bm q}^{\left(- \right)}$ 
modes correspond to the acoustic and optical PS waves, respectively. 
The energy dispersions for the PS waves are shown in Fig.~\ref{fig:sw} (a). 
The acoustic mode softens around the phase boundary and $\epsilon_{{\bm q}={\bm 0}}^{\left( + \right)}$ 
becomes zero at the boundary. 

Through the systematic analyses of the EHM, VtM and PSM, we obtain a whole picture for the collective charge dynamics in the DM insulating system. 
The low-energy excitations in $\alpha_x(\omega)$ and $\alpha_y(\omega)$ are identified as the acoustic and optical PS waves, respectively.  
The finite-excitation energy for the peak 
A$_{+}$ observed in $\alpha_x(\omega)$ at the phase boundary (see Fig.~\ref{fig:EHM} and Fig.~\ref{fig:vt}) is due to the finite size effect, and the energy of the acoustic mode becomes zero at the phase boundary. 
This soft collective mode is connected to the electric-polarization flip in the CO phase and the intra-dimer bonding-antibonding excitation, so-called the dimer-excitation, in the DM phase. 
The results are summarized in Fig.~\ref{fig:sw}(b).

Finally, the recent experimental results are interpreted from the view point of the present collective charge excitation. 
The optical conductivity spectra have been measured in $\kappa$-(ET)$_2$Cu$_2$(CN)$_3$\cite{kezsmarki,nakaya}, and other $\kappa$-type ET compounds~\cite{sasaki,faltermeier,dumm}, and studied in the theoretical viewpoints.~\cite{gomi,yonemitsu} 
Characteristic three-peak structures are commonly observed in the mid-infrared region around 0.1-0.4eV, where the highest and second-highest peaks are called the dimer and Hubbard excitations, respectively. 
The lowest peak is sensitive to the band width, and is not assigned yet.  
As shown in Fig.~\ref{fig:EHM}, the optical absorption peaks in the present theory are classified into the inter-MO excitations (peaks C and D), 
the Hubbard excitations (peaks B and E), 
and the intra-dimer excitation connected to the collective excitation (peak A). 
In the case where the DM and CO phases compete with each other, as examined in the present paper and expected in $\kappa$-(ET)$_2$Cu$_2$(CN)$_3$,  excitation energies for the peaks C and D are higher than that for the peak A. 
A possible interpretation for the experimental spectra is that the highest and second-highest peaks in the mid-infrared region correspond to the inter-MO excitations and the Hubbard excitations, respectively. 
Strong polarization dependence seen in the peak A  may rule out the interpretation that the lowest peak, which shows a weak polarization dependence, is attributed to the intra-dimer excitation. 
In the materials which are far from the DM-CO phase boundary, the intra-dimer excitations are hybridyzed strongly with the inter-MO ones and may contribute to the spectra around the mid-infrared region. 

Recently, Itoh and coworkers find out a new peak in the terahertz region around 4meV in $\kappa$-(ET)$_2$Cu$_2$(CN)$_3$\cite{itoh}. This peak is only observed in the $y$ polarization. 
With decreasing temperature, the peak intensity increases markedly, in contrast to other peaks. 
This temperature dependence is strongly correlated with the remarkable increasing in the dielectric constant.~\cite{jawad} 
This newly finding peak in the terahertz region is a plausible candidate of the collective charge excitation. 
It is supposed that $\kappa$-(ET)$_2$Cu$_2$(CN)$_3$ is located near the phase boundary between the DM and polar-CO phases, and below 25K where the dielectric constant takes a peak, inhomogeneous small polar-CO domains grow up. 
The increasing of the terahertz-peak intensity is attributed to development of the polar-CO region in low temperatures. 
We further expect a coupling between the collective excitation and spins through the ${\bm S} \cdot {\bm S} Q^{x} Q^{x}$-type interaction in ${\cal H}_J$~\cite{naka_kappa}, which gives rise to the competitive relation between the polar CO and spin correlation. When the observed terahertz peak is attributed to the collective excitation in the polar CO region, it is supposed that increasing of the spin fluctuation shifts the system to the phase boundary and reduces the collective-mode energy. 
More detailed theoretical studies including inhomogeneous effects are required to 
confirm this scenario. 
The direct observations of the collective mode by the electron-energy loss spectroscopy, or the inelastic x-ray scattering are helpful to a comprehensive understanding of the charge dynamics. 

We thank T.~Sasaki, S.~Iwai, and J.~Nasu for helpful discussions.
This work was supported in part by Grant-in-Aid for Scientific Research Priority Area from the Ministry of Education, Science and Culture of Japan.
MN is financially supported as a Research Fellow of JSPS. 




\begin{thebibliography}{99}
\bibitem{exchange}
Exchange constants in the PSM are given by 
$J_{ij}^{\gamma \gamma} = - ( {\tau_{ij}^{\gamma \bar{\gamma}}}^{2} + {\tau_{ij}^{\bar{\gamma} \gamma}}^{2})/\Delta_{\gamma \gamma}$, 
$J_{ij}^{\gamma \bar{\gamma}} = - ( {\tau_{ij}^{\gamma \gamma}}^{2} + {\tau_{ij}^{\bar{\gamma} \bar{\gamma}}}^{2} )/ \Delta_{\gamma \bar{\gamma}}$, 
$K_{ij}^{\gamma \gamma} = \tau_{ij}^{\gamma \bar{\gamma}} \tau_{ij}^{\bar{\gamma} \gamma} ( \Delta_{\gamma \gamma}^{-1} + \Delta_{\bar{\gamma} \bar{\gamma}}^{-1} )$, 
$K_{ij}^{\gamma \bar{\gamma}} = 2 \tau_{ij}^{\gamma \gamma} \tau_{ij}^{\bar{\gamma} \bar{\gamma}} / \Delta_{\gamma \bar{\gamma}}$, 
$I_{ij}^{x \pm} = \pm ( \tau_{ij}^{- +} \tau_{ij}^{+ +} - \tau_{ij}^{+ -} \tau_{ij}^{- -} ) ( \Delta_{\pm \pm}^{-1} + \Delta_{+ -}^{-1} )$, 
$I_{ij}^{\pm x} = \pm ( \tau_{ij}^{+ -} \tau_{ij}^{+ +} - \tau_{ij}^{- +} \tau_{ij}^{- -} ) ( \Delta_{\pm \pm}^{-1} + \Delta_{+ -}^{-1} )$, 
where 
$\bar{\gamma} = ( -, + )$ for $\gamma = ( +, - )$. 
We define $\tau_{ij}^{\gamma \gamma'}=\sum_{\mu \mu'} U_{\gamma \mu}^{\dagger} t_{ij}^{\mu \mu'}U_{\mu' \gamma'} $,@ 
$\Delta_{+ +} = V_{A} - 2 t_{A}$, 
$\Delta_{- -}=V_{A} + 2 t_{A}$ and 
$\Delta_{+ -}=\Delta_{- +}=V_{A}$.
\bibitem{sw}
We define 
$J_{A {\bm q}} = 2t_{A} - J_{z}^{B} - 2 J_{z}^{p} - J_{zz}^{B} - 2 J_{zz}^{p} - (V_{B}/2 - 2 J_{+-}^{B}) \cos {\bm q} \cdot {\bm a_{x}}$, 
$J_{B {\bm q}} = [ - (V_{p} - V_{q})/2 + 2 J_{+-}^{p} ] [ \cos ({\bm q}/2) \cdot ({\bm a}_{x} - {\bm a}_{y}) + \cos ({\bm q}/2) \cdot ({\bm a}_{x} + {\bm a}_{y}) ]$, 
$J_{C {\bm q}} = ( - V_{B}/2 + 2 J_{++}^{B} ) \cos {\bm q} \cdot {\bm a}_{x}$ and 
$J_{D {\bm q}} = [ - (V_{p} - V_{q})/2 + 2 J_{++}^{p} ] [ \cos ({\bm q}/2) \cdot ({\bm a}_{x} - {\bm a}_{y}) + \cos ({\bm q}/2) \cdot ({\bm a}_{x} + {\bm a}_{y}) ]$, 
where 
$J_{zz}^{B} = - ( t_{B}^{2}/2 ) ( \Delta_{- -}^{-1} + \Delta_{+ +}^{-1} - 2 \Delta_{- +}^{-1} )$, 
$J_{z}^{B} = - ( t_{B}^{2}/2 ) ( \Delta_{- -}^{-1} - \Delta_{+ +}^{-1} )$, 
$J_{++}^{B} = - ( t_{B}^{2}/4 ) ( \Delta_{- -}^{-1} + \Delta_{+ +}^{-1} )$, 
$J_{+-}^{B} = - ( t_{B}^{2}/2 ) \Delta_{- +}^{-1}$, 
$J_{zz}^{p} = - ( t_{p}^{2}/2 + t_{q}^{2}/2 ) ( \Delta_{- -}^{-1} + \Delta_{+ +}^{-1} - 2 \Delta_{- +}^{-1} )$, 
$J_{z}^{p} = - ( t_{p}^{2}/2 + t_{q}^{2}/2 ) ( \Delta_{- -}^{-1} - \Delta_{+ +}^{-1} )$, 
$J_{++}^{p} = - ( t_{p}^{2}/4 - t_{q}^{2}/4 ) ( \Delta_{- -}^{-1} + \Delta_{+ +}^{-1} )$ and  
$J_{+-}^{p} = - ( t_{p}^{2}/2 - t_{q}^{2}/2 ) \Delta_{- +}^{-1}$.
\end{thebibliography}
\end{document}